\documentclass[12pt]{amsart}
\usepackage{amssymb}
\usepackage{amsmath,latexsym}
\newcommand{\R}{\mathbb{R}}
\newcommand{\N}{\mathbb{N}}

\newcommand{\C}{\mathbb{C}}
\newcommand{\Z}{\mathbb{Z}}
\font\eufm=eufm10
\def\frak#1{\hbox{\eufm#1}}
\newcommand{\bd}{\begin{document}}
\newcommand{\ed}{\end{document}}
\newcommand{\be}{\begin{enumerate}}
\newcommand{\ee}{\end{enumerate}}
\newcommand{\bi}{\begin{itemize}}
\newcommand{\ei}{\end{itemize}}
\newcommand{\ba}{\begin{array}}
\newcommand{\ea}{\end{array}}
\newcommand{\vs}{\vspace*{0.3\baselineskip}}
\newcommand{\vsm}{\vspace*{-0.3\baselineskip}}
\newcommand{\kom}[1]{{\em #1}\newline}
\newtheorem{defi}{Definition}[section]
\newtheorem{tw}[defi]{Theorem}
\newtheorem{prop}[defi]{Proposition}
\newtheorem{lem}[defi]{Lemma}
\newtheorem{re}[defi]{Remark}
\newtheorem{col}[defi]{Corollary}
\newtheorem{ex}[defi]{Examples}
\newtheorem{zad}{Exercise}[section]
\newtheorem{zal}{Assumptions}[section]
\newtheorem{assumpt}[defi]{Assumptions}
\newcommand{\Om}{\Omega}
\newcommand{\om}{\omega}
\newcommand{\G}{\Gamma}
\newcommand{\D}{\Delta}
\renewcommand{\d}{\delta}
\newcommand{\ga}{\gamma}
\newcommand{\eps}{\epsilon}
\newcommand{\dg}{\dagger}
\newcommand{\ove}{\overline}
\newcommand{\ms}{\oplus}
\newcommand{\mt}{\otimes}
\newcommand{\dz}{\wedge}
\newcommand{\lra}{\longrightarrow}
\newcommand{\sign}{\mbox{$ sgn $}}
\newcommand{\rel}{\mbox{$\,$\rule[0.5ex]{1.1em}{0.2pt}$\triangleright\,$}}
\newcommand{\dow}{\hspace*{\fill}\rule{1.6ex}{1.6ex}\hspace*{1em}}
\newcommand{\dowl}{\hspace*{\fill}\rule{1ex}{1ex}\hspace*{1em}}
\newcommand{\sd}{\hspace{0.3ex}\tiny{\rhd\mbox{\hspace{-2ex}}<}\hspace{0.3ex}}
\newcommand{\mmt}[2]{\mbox{$\vphantom{}_{#1}\times_{#2}$}}
\newcommand{\gotg}{\frak g}
\newcommand{\gota}{\frak a}
\newcommand{\gotb}{\frak b}
\newcommand{\gotc}{\frak c}
\newcommand{\gothh}{\frak h}
\newcommand{\gott}{\frak t}
\newcommand{\hd}{\hat{\d}}
\newcommand{\oml}{\Omega_L^{1/2}}
\newcommand{\omr}{\Omega_R^{1/2}}
\newcommand{\omh}{\Omega^{1/2}}
\newcommand{\lo}{\lambda_0}
\newcommand{\ro}{\rho_0}
\newcommand{\lma}{\Lambda^{max}}
\newcommand{\timh}{\times_h}
\newcommand{\Gd}{\G^{(2)}}
\newcommand{\el}{e_L}
\newcommand{\er}{e_R}
\newcommand{\GG}{\G_1\times\G_2}
\newcommand{\gdot}{\hspace{-0.1em}\cdot\hspace{-0.1em}}
\newcommand{\tran}{\frown\hspace{-2.2ex}|\hspace{1.9ex}}
\newcommand{\la}[2]{\Lambda_{#1#2}}
\newcommand{\kad}{ad^{\#}}
\newcommand{\wl}[1]{\vphantom{X}_{#1}{\G}}
\newcommand{\te}{\tilde{e}}
\newcommand{\notka}[1]{\mbox{\newline\framebox{#1}\newline}}
\newcommand{\sA}{\mbox{$\mathcal A$}}
\newcommand{\sT}{\mbox{$\mathcal T$}}
\newcommand{\sB}{\mbox{$\mathcal B$}}
\newcommand{\sF}{\mbox{$\mathcal F$}}
\newcommand{\sO}{\mbox{$\mathcal O$}}
\newcommand{\sD}{\mbox{$\mathcal D$}}
\newcommand{\sS}{\mbox{$\mathcal S$}}
\newcommand{\sY}{\mbox{$\mathcal Y$}}
\newcommand{\sH}{\mbox{$\mathcal H$}}
\newcommand{\sK}{\mbox{$\mathcal K$}}
\newcommand{\sL}{\mbox{$\mathcal L$}}
\newcommand{\hY}{\mbox{$\hat{Y}$}}
\newcommand{\hS}{\mbox{$\hat{S}$}}
\newcommand{\hX}{\mbox{$\hat{X}$}}

\newcommand{\dif}{differential }
\newcommand{\gru}{groupoid }
\newcommand{\grus}{groupoids }
\newcommand{\ti}{\tilde}
\newcommand{\halden}{half density }
\newcommand{\haldens}{half densities }
\renewcommand{\top}{topological }
\newcommand{\Setrel}{\mbox{\rm SetRel}}
\newcommand{\cstardwa}{\mbox{$C^*_r(\Gamma\times\Gamma)$}}
\newcommand{\cred}{\mbox{$C^*_r$}}
\def\tgr{{\bf t}}
\def\sgr{{\bf s}}
\def\fgr{{\bf f}}
\def\xgr{{\bf x} }
\def\ygr{{\bf y} }
\def\kgr{{\bf k} }
\def\ngr{{\bf n} }
\def\zgr{{\bf z} }
\def\rhogr{{\boldsymbol \rho}}
\def\Ygr{{\bf Y}}

\newcommand{\ekin}{{\rm T}}
\newcommand{\inter}{{\rm H}_{int}}
\newcommand{\ham}{{\rm H}}
\newcommand{\skal}{\,|\,}
\newcommand{\tr}{\mathrm{Tr}}
\newcommand{\bound}{\mathcal{B}}
\newcommand{\cj}{\rm K_J}
\begin{document}
\title[Bogolyubov inequality for the ground state and its application \dots]
{Bogolyubov inequality for the ground state and its application 
 to  interacting rotor systems.}
\author{Jacek Wojtkiewicz, Wies{\l}aw Pusz}
\address{Department for Mathematical Methods in Physics
\\ Faculty of Physics, Warsaw University\\
  Pasteura 5, 02-093 Warszawa, Poland\\
e-mail: ${\rm wjacek@fuw.edu.pl}$ (J.W.), ${\rm wpusz@fuw.edu.pl}$ (W.P.)}
\author{Piotr Stachura}
\address{The Faculty of Applied Informatics and Mathematics\\
Warsaw University of Life Sciences-SGGW\\
ul. Nowoursynowska 159, 02-776 Warszawa, Poland\\
e-mail: ${\rm piotr\_{}stachura1@sggw.pl}$}
\begin{abstract}
We have formulated and proved the Bogolyubov inequality for operators at zero temperature.
So far this inequality has been known for matrices, and we were able to extend it to certain class of operators. We have also 
applied this inequality to the system of interacting rotors. We have shown that if: {\em i)} the dimension of the
lattice is 1 or 2, {\em ii)} the interaction decreases  sufficiently fast  with  a distance,  
and {\em iii)} there is an energy gap over the ground state, then the spontaneous
magnetization in the ground state is zero, i.e. there is no LRO in the system.
 We present also heuristic arguments (of perturbation-theoretic nature)
 suggesting that one- and two-dimensional system of interacting rotors
 has the energy gap independent of the system size
 if the interaction is sufficiently small. 
\end{abstract}
\maketitle
\section{Introduction}
The {\em  Mermin -- Wagner theorem}  \cite{MW} is one of the most important results, 
concerning phase transitions or their absence in systems described by statistical mechanics. In short, 
it claims that {\em in one- and two-dimensional quantum lattice systems, possessing continuous symmetry, and with short-range
interactions, there is no
Long-Range Order (LRO) in non-zero temperature}. 

The Mermin-Wagner theorem (M-W theorem) has been generalized in many directions. 
Some of the most important results are:  no  LRO in boson systems \cite{Hohenberg}; extension of the theorem to classical models \cite{Mermin67}; 
spatial fall-off of correlation functions \cite{FisherJasnow}; lack of magnetic and superconducting orderings in   itinerant  fermion systems 
(Hubbard-like models) \cite{Ghosh}, \cite{KomaTasaki}; quantum rotor systems \cite{PaChor}; no LRO in spin systems
with lack of translational symmetry \cite{Cassi}. For a review, see \cite{GelfertNolting2001}.

The original M-W theorem  deals only with positive temperature but, under certain additional assumptions, it has also been generalized to 
the {\em zero temperature} \cite{Auerbach}\cite{SSZ}.
It turns out that the ground state of one- or two-dimensional
systems with continuous symmetry
group can be ordered in some cases (XY or Heisenberg models on square lattice -- see \cite{NevesPerez}),
whereas  other models possess disordered ground state ('Resonating Valence Bond models' -- see \cite{AKLT}).
 One possible source of a disorder at zero temperature is  {\em an energy gap}
between the ground state and excited states: if a given system is gapped, then it is disordered also in the ground state.
 This  has been shown for the lattice spin systems \cite{Auerbach} and 
itinerant fermion systems \cite{SSZ}, \cite{SuSuzuki},  \cite{Noce}.

The M-W theorem has been proved with the use of {\em Bogolyubov inequality} \cite{Bogolubow62}.
(Later on, there appeared also another techniques of proving lack of LRO in one- and two-dimensional systems
with continuous symmetry group for positive temperatures; for an excellent presentation, see  \cite{Simon93}).
The Bogolyubov inequality is also the basic technical tool in extensions of M-W theorems.
An important aspect here is the dimension of the Hilbert space on a single site. If this dimension is finite,
then the Bogolyubov inequality is a matrix inequality and its proof is relatively easy. However, when this dimension
is infinite, then certain operator-theoretic considerations enter the game. Such an infinite-dimensional 
version of the Bogolyubov inequality (for positive temperature) has been proved in  \cite{BouzianeMartin} {
and applied in an analogon of the M-W theorem for interacting rotors in \cite{PaChor}.

 However -- to our best knowledge -- the question of infinite dimensional version of the Bogolyubov inequality
for {\em zero temperature} is still open. It seems that the same concerns the zero-temperature M-W 
theorem for systems with infinite dimensional Hilbert space. This opportunity motivated us to undertake 
efforts on these areas. In the present  paper, we describe the results obtained. They can be summarized as follows: 
\begin{itemize}
\item[{\em i)}] We have formulated and proved an infinite-dimensional version of the Bogolyubov inequality for
zero temperature.
\item[{\em ii)}] We have applied it to the system of interacting rotors and  
 have shown that if there exists an energy gap over the ground state, then there is no LRO  at zero temperature.
\end{itemize} 
We also present (non-rigorous) arguments  
for existence of the gap if the interaction between rotors is sufficiently small. 

The organization of the paper is as follows. In the next section we formulate and prove the zero-temperature version of the Bogolyubov inequality for certain class of operators.   
In the third section, we define Hamiltonians for  interacting rotor systems, the magnetization as the measure of ordering, and prove the zero-temperature Mermin-Wagner
theorem for these systems. 
We consider the planar and spherical rotors. We present also arguments for existence of energy gap
in the weakly interacting rotor systems; they are based on perturbation theory. The fourth section 
is devoted to the summary and description of some open problems. The Appendix  contains conventions and formulas
 for Legendre Polynomials and spherical harmonics we use.

\section{Zero temperature Bogoliubov inequality}
\label{sec:BITzero}
\newcommand{\OP}{Op}

\subsection{Finite dimensional Bogolyubov inequalities}
We begin by recalling some well known facts concerning the original Bogolyubov inequality \cite{MW}.
Let $\sH$ be a {\em finite dimensional} Hilbert space and $\ham$ a selfadjoint operator on $\sH$ (the Hamiltonian of the system). Define the function \cite{MW}
$(\cdot\, ,\, \cdot ):\bound(\sH)\times\bound(\sH)\rightarrow \C$:
\begin{equation}\label{duhamel}
(A,B) : = \frac{1}{\tr (e^{-\beta \ham})} \sum_{i,j}\hspace{-0.1em}{}^{'}\, (j\skal A^\dagger i) \, (i\skal Bj)
\frac{e^{-\beta E_i} - e^{-\beta E_j} }{E_j - E_i},
\end{equation}

where the summation is performed over all eigenstates of $\ham$ and  $E_i$ is an eigenenergy corresponding to the $i$-th eigenstate.
The prime at the sum sign denote that the summation is performed excluding pairs with the same energy.

\noindent {\bf Remark.} The function (\ref{duhamel}) is closely related to the {\em Duhamel function}, see  \cite{Simon93}, \cite{DLS} 

One checks that $(A,B)$ is a hermitean, sesquilinear, non-negative form  on $\bound(\sH)$.  Therefore the Schwarz inequality holds:
\[
|(A,B)|^2\leq (A, A) (B, B).
\]
Using this formula one derives  the {\em Bogolyubov inequality}:
\begin{equation}
|\langle [A,B]\rangle|^2 \leq \frac{\beta}{2}\langle[[B,H],B^\dg]\rangle\langle A^\dg A + A A^\dg\rangle
\label{originalBI}
\end{equation}
where $\langle\,\cdot\,\rangle$ denote standard thermal average.
The definition of $(A,B)$ and the Bogolyubov inequality are meaningful  for positive temperatures. 
However, one can
modify them to the situation of zero temperature ($\beta=\infty$)
\cite{SSZ}. To this aim let us define:
\begin{equation}
(A,B): = \frac{1}{Z} \sum_{m}\sum_\alpha
\frac{(\alpha \skal  A^\dg  m ) (  m \skal  B \alpha )+
( \alpha \skal  B  m ) (  m \skal A^\dg \alpha ) }{E_m-E_0}
\label{IlSkalSSZ}
\end{equation}
where $\alpha$ is an index for  ground states, $Z =\sum_\alpha (  \alpha \skal  \alpha )$, $E_0$ is the ground state energy and  $m$ refers to excited states with eigenenergy $E_m$. 
As before,   (\ref{IlSkalSSZ}) defines a hermitean, sesquilinear, non-negative formt on $\bound(\sH)$. Using the Schwarz inequality 
one derives  the {\em zero-temperature version of the Bogolyubov inequality}:
\begin{equation}
|\langle [A,B]\rangle_0|^2 \leq \frac{1}{\Delta}\langle[B,[H,B^\dg]]\rangle_0\langle A^\dg A + A A^\dg\rangle_0
\label{originalBI}
\end{equation}
where the average is taken over the ground states of the system and $\Delta:=E_1-E_0$. (See also \cite{Auerbach},  \cite{Noce} for related considerations.) 
Our goal now is  to extend the definition of $(A,B)$ and the Bogolyubov inequality for $A,B$ being operators on  an infinite dimensional Hilbert space. 

\subsection{Infinite dimensional, zero temperature Bogolyubov inequality.}
Let  $\ham$ be a  selfadjoint, bounded from below operator with compact resolvent acting on a separable Hilbert space $\sH$ and let $\displaystyle \ham =\sum_{n=0}^\infty E_n P_n$ be  its 
spectral decomposition. Thus we have:
\begin{equation}\label{spectral}\begin{split}
&E_0<E_1<\dots <E_n <\dots\quad,\quad \lim_{n\rightarrow\infty}E_n=\infty\\
&P_n^*P_m=\delta_{nm}P_n\quad,\quad \sum_n P_n=I\quad,\quad  N_n:=\dim P_n(\sH) <\infty.
\end{split}
\end{equation}
For further considerations we shall fix  an  $\ham$-invariant subspace $\sS_\ham\subset \sH$   such that $P_n(\sH)\subset \sS_\ham$ for any $n=0,1,\dots$. 
Clearly  $\sS_\ham$ is dense in $\sH$.  Therefore any linear(possibly unbounded)  operator  $C:\sS_\ham\rightarrow \sH\quad(C\in\sL (\sS_\ham,\sH))$  has a well defined adjoint operator.

In what follows we shall consider the following set of operators
\begin{equation}
\OP(\sS_\ham):=\{ C\in \sL (\sS_\ham,\sH) :\sS_\ham\subset D(C^*)\}
\end{equation} 
Notice that for any $C\in \OP(\sS_\ham)$  the operator $C^*$ is densely defined,  so $C$ is closeable and $\overline{C}=C^{**}$.
Let   $C^\dagger:=C^*|_{\sS_\ham}$. Moreover $C^\dagger \in\OP(\sS_\ham)$; in fact,  since $C^\dagger\subset C^*$,  we have 
$\overline{C}=C^{**}\subset (C^\dagger)^*$ i.e. $ C\subset \overline{C}\subset(C^\dagger)^*$. Now,  
$(C^\dagger)^\dagger=(C^\dagger)^*|_{\sS_\ham}=C$ and the map $\OP(\sS_\ham)\ni C\mapsto C^\dagger \in \OP(\sS_\ham)$ is an {\em antylinear involution.} 

A  {\em  bounded} operator $A$ acting on  $\sH$   defines an element of $\OP(\sS_\ham)$ by restriction to $\sS_\ham$.
On the other hand $A=\overline{(A|_{\sS_\ham})}$. In this sense  we  shall  write $\bound(\sH)\subset\OP(\sS_\ham)$.
Clearly $\sS_\ham$ is an essential domain for $\ham$ and for   $\ham_0:=\ham|_{\sS_\ham}$ we have $\ham_0\in \OP(\sS_\ham)$ and $\ham_0^\dagger=\ham_0$. 
Notice that $\OP(\sS_\ham)$ is a linear space. For  any  $C\in \OP(\sS_\ham)$ and any $n=0,1,\dots$ operators $ CP_n\,\,{\rm and}\,\,C^* P_n=C^\dagger P_n$ 
are well defined  and have finite dimensional rank.
\begin{lem}\label{lem1} a) Let  $A\in \bound(\sH)$ and $A(\sS_\ham)\subset \sS_\ham $. Then   $CA, A^*C\in \OP(\sS_\ham)$ for any $C\in \OP(\sS_\ham)$.\\
b) Let $C, B\in \OP(\sS_\ham)$ such that  $C(\sS_\ham), B(\sS_\ham)\subset \sS_\ham$.  Then $C^\dagger B\in \OP(\sS_\ham)$.\\
c) Let  $C\in \OP(\sS_\ham)$ such that  $C(\sS_\ham), C^\dagger (\sS_\ham)\subset \sS_\ham$.  Then $[C^\dagger, \ham_0],\, [C, \ham_0]\in \OP(\sS_\ham)$.
\end{lem}
\noindent
{\em Proof:} Ad a) Clearly $C A, A^*C\in \sL (\sS_\ham,\sH)$. Moreover $\sS_\ham\subset D((C A)^*)$ due to the inclusion $(C A)^*\supset A^* C^*$. Similarly,
$(A^*C)^*\supset C^* A$ and $\sS_\ham$ is $A$ invariant,  therefore  $\sS_\ham\subset D((A^*C)^*)$.\\
Ad b) By $B$ invariance of $\sS_\ham$: $C^\dagger B:\sS_\ham\rightarrow \sH$. Since  $(C^\dagger B)^*\supset B^* (C^\dagger)^*\supset B^* C$ 
and  $\sS_\ham$ is $C$-invariant,  the domain of $B^* C$ contains $\sS_\ham$, so $C^\dagger B\in \OP(\sS_\ham)$. \\
Ad c) Since   $[C^\dagger, \ham_0]=C^\dagger \ham_0-\ham_0 C^\dagger\in  \sL (\sS_\ham,\sH)$ this assertion follows from b) and the fact that $\OP(\sS_\ham)$ is a linear space.\\\dowl

Let us define a family of sesquilinear forms $\rho_{nk}$ on $\OP(\sS_\ham)$ by:
\begin{equation}\label{def-omega-ro}
\rho_{nk}(C,B):=\tr(P_k C^* P_n B P_k)=\sum_{j=1}^{N_k}(Ce_j\skal P_n B e_j)
\,,\quad n,k =0,1,2,\dots
\end{equation}
where $(e_j)$ is an o.n basis in $P_k(\sH)$ (and $N_k:=\dim P_k(\sH)$).
\begin{lem}\label{rho-prop}
\begin{enumerate}
\item $\rho_{nk}$ are non-negative, hermitean forms on $\OP(\sS_\ham)$,
\item  $\rho_{nk}(C,B)=\rho_{kn}(B^\dagger,C^\dagger)$ for any  $C,B\in\OP(\sS_\ham)$
\item  Let $C, B\in \OP(\sS_\ham)$ such that  $C(\sS_\ham), C^\dagger(\sS_\ham)\subset \sS_\ham$. Then 
$$\rho_{nk}(B,[C,\ham_0])=(E_k-E_n) \rho_{nk}(B,C).$$
\item  For any $C, B\in \OP(\sS_\ham)$ the series $\displaystyle \sum_n \rho_{nk}(C,B)$ is absolutely convergent and
$$\displaystyle \sum_n \rho_{nk}(C,B)=\sum_{j=1}^{N_k}(Ce_j\skal  B e_j),$$
where  $(e_j)$ is o.n. basis in $P_k(\sH)$.
\end{enumerate}
\end{lem}
\noindent
{\em Proof:} Statement (1) is  a direct consequence of the definition;.  The statement (2) follows from calculation:
\begin{equation*}
\begin{split}
\rho_{nk}(C,B)&=\tr(P_k C^*P_nBP_k)=\tr(P_n B P_k C^* P_n)=\tr(P_n B P_k C^\dagger P_n)=\tr(P_n (B^\dagger)^* P_k C^\dagger P_n)\\
&=\rho_{kn}(B^\dagger,C^\dagger).
\end{split}
\end{equation*}

To prove the statement (3) let us observe that  by lemma \ref{lem1}(c) $[C,\ham_0]\in\OP(\sS_\ham)$.  Since $\ham_0 P_k=E_k P_k$ and $P_n \ham_0 C P_k=E_n P_n C P_k$ we have:
\begin{equation*}
\begin{split} \rho_{nk}(B,[C,\ham_0])& =\tr(P_k B^*P_n (C \ham_0-\ham_0 C)P_k)=\tr(P_kB^*P_nC \ham_0 P_k-P_k B^* P_n\ham_0 C P_k)=\\
&=\tr(E_k P_k B^* P_nC P_k-E_n P_k B^* P_n C  P_k))=(E_k-E_n)\tr( P_k B^* P_n C P_k)\\&=(E_k-E_n)\rho_{nk}(B,C).\end{split}
\end{equation*}

Let us prove (4). For any $C\in \OP(\sS_\ham)$ the positive series  $\sum_n \rho_{nk}(C,C)$ is convergent:
\begin{equation*}
\begin{split}
 \infty >\sum_{j=1}^{N_k}(C e_j\skal C e_j)&=\sum_{j=1}^{N_k}\sum_n(C e_j\skal P_n C e_j)= \sum_n \sum_{j=1}^{N_k}(C e_j\skal P_n C e_j)=\\
&=\sum_n \tr( P_kC^* P_n C P_k)=\sum_n \rho_{nk}(C,C).
\end{split}\end{equation*}
Now the assertion follows from the polarization formula for $\rho_{nk}(C,B)$.
\dowl

\noindent Let us define:
\begin{equation}\label{ilosk}
\OP(\sS_\ham)\times \OP(\sS_\ham)\ni (B,C)\mapsto (B,C)_0:=\sum_{n >0 }\frac1{E_n-E_0}(\rho_{n0}+\rho_{0n})(B,C)
\end{equation}
Notice that the sequence $a_ n:=\frac{1}{E_n-E_0}\,,\,n\neq 0$ is bounded  so by (2) and (4) of the  lemma \ref{rho-prop}
 the  series in (\ref{ilosk})  is absolutely convergent and the definition makes sense.

\begin{prop} \label{ilosk-prop} The form $(\cdot\,,\,\cdot)_0$ is sesquilinear, hermitean and non-negative. \\
 Moreover for  any $B,C\in\OP(\sS_\ham)$: 
\begin{enumerate}
\item $\displaystyle (B,C)_0=(C^\dagger,B^\dagger)_0\,;$
\item  $\displaystyle (B, [C,\ham_0])_0=\tr(P_0[C, B^\dagger]P_0)\,$, whenever $B^\dagger(\sS_\ham),C^\dagger(\sS_\ham),C(\sS_\ham)\subset \sS_\ham$.
\end{enumerate}
\end{prop}
{\em Proof:} The form $(\cdot\,,\,\cdot)_0$ is a sum of hermitean, non-negative forms. The  formula (1) follows from statement (2)  of the Lemma \ref{rho-prop}. 
By the  item (3) Lemma \ref{rho-prop}:
$$\frac1{E_n-E_0}(\rho_{n0}+\rho_{0n})(B, [C,\ham_0])=-\rho_{n0}(B,C)+\rho_{0n}(B,C)=\rho_{n0}(C^\dagger, B^\dagger)-\rho_{n0}(B,C).$$
Now, using the  item (4) of the same lemma:
\begin{equation*}
\begin{split}
\sum_n\left[\rho_{n0}(C^\dagger, B^\dagger)-\rho_{n0}(B,C)\right]&=\sum_{j=1}^{N_0}\left[(C^\dagger e_j \skal B^\dagger e_j)-(B  e_j \skal C e_j)\right]=
\sum_{j=1}^{N_0}\left[(e_j \skal  C B^\dagger e_j)-(  e_j \skal B^\dagger C e_j)\right]=\\
&=\tr(P_0[C,B^\dagger]P_0),
\end{split}\end{equation*}
where the last but one equality follows from the inclusion $B^\dagger(\sS_\ham)\subset \sS_\ham$.
\dowl

\begin{col}\label{col1}
Let $B, C\in\OP(\sS_\ham)$ such that $C(\sS_\ham), C^\dagger (\sS_\ham)\subset \sS_\ham$. Then
\begin{equation*}
|(B, [C^\dagger, \ham_0])_0|^2\leq (B,B)_0 ([C^\dagger, \ham_0], [C^\dagger, \ham_0])_0
\end{equation*}
\end{col}
{\em Proof:} This is the Schwarz inequality for $(\cdot\,,\,\cdot)_0$  and operators $B$ and $[C^\dagger, \ham_0]$ (cf. statement (c) of Lemma \ref{lem1}).
\\\dowl

Let us observe that  $[C^\dagger, \ham_0]^\dagger=[\ham_0,C]$, so  by (2) of Prop.\ref{ilosk-prop} we have
$$([C^\dagger, \ham_0],[C^\dagger, \ham_0])_0=\tr(P_0[C^\dagger,[\ham_0,C]]P_0,)$$ 
Now Corollary \ref{col1} reads:
$$|(B, [C^\dagger, \ham_0])_0|^2\leq (B,B)_0 \tr(P_0[C^\dagger,[\ham_0,C]]P_0)$$ 
If in addition $B^\dagger(\sS_\ham)\subset \sS_\ham$, by the same argument as above, we obtain:
$$(B, [C^\dagger, \ham_0])_0=\tr(P_0[C^\dagger,B^\dagger]P_0).$$
This way we have proved:
\begin{prop}\label{prop-bogol} Let  $B,C\in\OP(\sS_\ham)$ be such that $B^\dagger(\sS_\ham), C(\sS_\ham),  C^\dagger (\sS_\ham)\subset \sS_\ham$. Then
\begin{equation}\label{eq:prop-bogol}
|\tr(P_0[C^\dagger,B^\dagger]P_0)|^2 \leq (B,B)_0 \tr(P_0[C^\dagger,[\ham_0,C]]P_0)
\end{equation}
\end{prop}\dowl

\noindent
Finally,  we can formulate our basic inequality.
\begin{prop} \label{prop-bogol-1} (``Bogolyubov inequality'') \\
Let $B,C\in\OP(\sS_\ham)$ be such that $B^\dagger(\sS_\ham), C(\sS_\ham),  C^\dagger (\sS_\ham)\subset \sS_\ham$. Then
\begin{equation}\label{bogolubow-k}
|\tr(P_0[C^\dagger,B^\dagger]P_0)|^2 \leq  \left(\frac{1}{E_{1}-E_0}\sum_{j=1}^{N_0}(||B e_j||^2+||B^\dagger e_j||^2)\right) \tr(P_0[C^\dagger,[\ham_0,C]]P_0),
\end{equation}
where $(e_j)$ is any o.n. basis in $P_0(\sH)$.
\end{prop}
{\em Proof:}
We have the estimate:
\begin{equation*}
\begin{split}
(B,B)_0=&\sum_{n>0}\frac1{E_n-E_0}(\rho_{n0}+\rho_{0n})(B,B) \leq  \frac{1}{E_{1}-E_0}\sum_{n > 0}\left[\rho_{n0}(B,B)+\rho_{0n}(B,B)\right]\\
\leq & \frac{1}{E_{1}-E_0}\sum_{n=0}^\infty\left[\rho_{n0}(B,B)+\rho_{0n}(B,B)\right] = \frac{1}{E_{1}-E_0}\sum_{n=0}^\infty\left[\rho_{n0}(B,B)+\rho_{n0}(B^\dagger,B^\dagger)\right]\\
=& \frac{1}{E_{1}-E_0}\sum_{j=1}^{N_0}\left(||Be_j||^2+||B^\dagger e_j||^2\right),
\end{split}
\end{equation*}
where the last equality follows from statement (4) of  the Lemma \ref{rho-prop}.
Inserting this estimate into the inequality in the Proposition \ref{prop-bogol} we get the result.\\\dowl


\section{Lack of ordering}
\label{sec:MWTzero}
\subsection{Definitions and basic properties of considered systems.}
Let  $\Lambda$  be a finite subset of the simple cubic lattice in $d$ dimensions:
 $\Lambda\subset \Z^d$.
We assume that  $\Lambda$ is a (discrete) hypercube and that {\em the number of sites along every edge is even}; let us fix $2N$ to be
the length of the hypercube edge:
\begin{equation}
\label{def-sieci}
\Lambda_N:=\{\xgr\in \Z^d: -N+1\leq x_i\leq N \,,\,i=1,\dots,d\}
\end{equation}
By  $|\Lambda_N|$ we denote  the number of sites in $\Lambda_N$ i.e. $|\Lambda_N|=(2N)^d$. With every site $\xgr\in\Lambda_N$ we associate  a separable, 
infinite dimensional Hilbert space $\sH_\xgr$ 
and the Hilbert space associated to the whole system is the tensor product $\displaystyle \sH_{\Lambda_N} := \mathop{\otimes}_{\xgr\in\Lambda_N}\sH_\xgr$. 
We will consider two systems: {\em planar and spherical rotors}.

{\em Planar rotors.} A position of a planar rotor at a site $\xgr$ is given by $\ngr_\xgr$ -- a unit vector in $\R^2$ or
equivalently by $\varphi_\xgr\in [0,2\pi[$; thus   $\sH_\xgr=L^2(S^1)$ and the total Hilbert space $\sH_{\Lambda_N}=L^2(S^1\times\dots\times S^1)$ 
is the Hilbert space of square integrable functions on $|\Lambda_N|$ dimensional torus $M$.

{\em Spherical rotors.} A position of a spherical rotor at a site $\xgr$ is given by $\ngr_\xgr$ -- a unit vector in $\R^3$ or
equivalently by angles $(\theta_\xgr,\varphi_\xgr)$; this time   $\sH_\xgr=L^2(S^2)$ and the total Hilbert space $\sH_{\Lambda_N}=L^2(S^2\times\dots\times S^2)$. In this case $M$ is a product of 
$|\Lambda_N|$ copies of $S^2$.

In both situations the hamiltonian of the system is a sum of three terms:  $\ekin$ -- kinetic energy, $V$ -- interaction of rotors on different sites and $\inter$ -- 
interaction with external field:
\begin{equation}\label{ham}
\ham:=\ekin +V+\inter
\end{equation}
The kinetic energy operator is given by $\ekin=-\frac{1}{2 I}\Delta$, for some positive constant $I$ (a moment of inertia), where $\Delta$   is the Laplace operator for the obvious 
riemmanian structures of a (flat) torus or  a product of spheres;  $V$ and $\inter$ are 
operators of multiplication by smooth functions. {\em Therefore  $\ham$ is
 an elliptic, second order differential operator on compact riemmanian  manifold $M$, formally selfadjoint on $C^\infty(M)$ and bounded from below }. It is known (\cite{Nicola})  that:
\begin{itemize}
\item $\ham$ is essentially selfadjoint and extends(uniquely) to a  bounded from below, selfadjoint operator (still denoted by $\ham$)  on $L^2(M)$;
\item the spectrum of $\ham$ consists of isolated eigenvalues with finite multiplicities;
\item eigenvectors of $\ham$ may be represented by  smooth functions.
\end{itemize}
Thus the spectral decomposition of $\ham$ is of the form given in (\ref{spectral}). As a space $\sS_\ham$ we take $C^\infty(M)$.

Our riemmanian manifolds $M$ are complete. In such a case it is known (see e.g. \cite{Davis}) that the {\em heat semigroup}
$e^{t \Delta }$ is an integral operator with a {\em strictly positive}  kernel (the heat kernel)  $K(t,x,y)$ on $]0,\infty[\times M\times M$. Therefore $e^{-t \ekin }$ is 
{\em positivity improving} and the same (by the Trotter product formula) is true for $e^{-t \ham}$. Therefore {\em the ground state is unique} 
i.e.  in the decomposition (\ref{spectral}) we have $\dim P_0(\sH)=1$ (see e.g.  Chapt. XIII of \cite{RS4} for detailes  or \cite{WPS1} for a very brief exposition).
By $$\om_0(A):=\tr(P_0 A P_0)$$  we shall denote the  expectation value in the ground state.

Let  $A,C\in\OP(C^\infty(M))$  and assume that $A,C, A^\dagger, C^\dagger$ preserve $C^\infty(M)$. Then inequalities (\ref{eq:prop-bogol}) and (\ref{bogolubow-k}) read:
\begin{align}
|\om_0([C^\dagger,A^\dagger])|^2 & \leq (A,A)_0 \,\om_0([C^\dagger,[\ham_0,C]])\label{bogol-1}\\
|\om_0([C^\dagger,A^\dagger)|^2 & \leq  \frac{\om_0(A^\dagger A + A A^\dagger)}{E_{1}-E_0} \,\om_0([C^\dagger,[\ham_0,C]])
\end{align}

\noindent
Let us specify in detailes our hamiltonians.\\ 
{\em Spherical Rotors. } The kinetic energy operator:
\begin{equation}\label{ekin-sph}
\ekin :=\frac{1}{2 I}\sum_\xgr L_\xgr^2 \quad,\,\,L_\xgr^2=-(\partial^2_{\theta_\xgr} +\cot \theta_{\xgr} \partial_{\theta_\xgr} +\frac1{\sin^2\theta_\xgr}\partial^2_{\varphi_\xgr}),
\end{equation}
where $I>0$ is the moment of inertia of rotors (we assume that all rotors have equal moments of inertia). The rotor-rotor interaction:
\begin{equation}\label{v-sph}
V:=\sum_{\xgr,\ygr} J_{\xgr\ygr}\sum_{m=-l}^{l}Y_l^m(\ngr_\xgr)\overline{Y_l^m(\ngr_\ygr})\quad\quad  J_{\xgr\ygr}:=J(|\xgr-\ygr|),
\end{equation}
where $Y_l^m$ are {\em spherical harmonics} (see Appendix for conventions we use). For the function $J: [0,\infty[ \rightarrow  [0,\infty[$ we assume:
\begin{equation}\label{zal-J}
\sum_{\xgr\in\Z^d}J(|\xgr|)|\xgr|^2=:\cj<\infty\,\,,\quad\quad  J(0)=0.
\end{equation}
The term describing the interaction with {\em an external magnetic field} $h\in\R$ is given by:
\begin{equation}\label{inter-sph}
\inter:=-h\sum_\xgr P_l(\cos\theta_\xgr),
\end{equation}
where $P_l$ denotes  the Legendre Polynomial (see Appendix). 

\noindent
{\em Planar Rotors. } The kinetic energy operator:
\begin{equation}\label{ekin-plan}
\ekin=-\frac{1}{2I}\sum_\xgr \frac{\partial^2}{\partial \varphi_\xgr^2},
\end{equation}
where $I>0$ is the moment of inertia of rotor. The rotor-rotor  interaction:
\begin{equation}\label{v-plan}
V=\sum_{\xgr, \ygr} J_{\xgr\ygr} \ngr_\xgr \cdot \ngr_\ygr
=\sum_{\xgr\ygr} J_{\xgr\ygr} (\cos\varphi_\xgr \cos\varphi_\ygr +
\sin\varphi_\xgr \sin\varphi_\ygr),
\end{equation}
the assumptions about $J_{\xgr\ygr}$ are as for spherical rotors  i.e. (\ref{zal-J}). The interaction term with an external field:
\begin{equation}\label{inter-plan}
\inter= - h \sum_\xgr \cos\varphi_\xgr
\end{equation}

\subsection{From Bogoliubov inequality to estimate for magnetization} In further consideration we shall focus on {\em magnetization} of the system as a measure of LRO. 
As is well known, this quantity is defined by:
\begin{equation}\label{mh}
m_N(h):=\frac{1}{|\Lambda_N|}\,\om_0\left(\sum_\xgr P_l(\cos\theta_\xgr)\right),
\end{equation}
for {\em spherical rotors}, and 
\begin{equation}\label{mh-planar}
m_N(h):=\frac{1}{|\Lambda_N|}\,\om_0\left(\sum_\xgr \cos\varphi_\xgr\right)
\end{equation}
for {\em planar} ones. Let us remark that according to our assumptions (cf (\ref{inter-sph},\ref{inter-plan})) the ground state $\om_0$ depends on $h$. 

We are going to use (\ref{bogol-1}) to get an estimate for magnetization.
Operator $A$ which appear in (\ref{bogol-1}) will be an operator of multiplication by smooth function (bounded, since our manifolds are compact) 
and  operator $C$  will be  first order differential operator with smooth coefficients. Clearly they belong to $\OP(C^\infty(M))$ and preserve  $C^\infty(M)$ (as do  their adjoints).
\subsubsection{Spherical Rotors}
Following the idea of   (\cite{PaChor}) we define operators:
\begin{align}\label{ak}
A_\kgr:= - \sum_\xgr e^{i \kgr \xgr} \cos\varphi_\xgr \sin\theta_\xgr  P_l'(\cos\theta_\xgr)\\
C_\kgr^\dagger:= \sum_\xgr e^{-i \kgr \xgr} L^+_\xgr= \sum_\xgr e^{-i \kgr \xgr} e^{i \varphi_\xgr} ( \partial_{\theta_\xgr} +i \cot \theta_\xgr  \partial_{\varphi_\xgr})
\end{align}
where $\kgr$ take values in the first Brillouin zone, i.e. 
\begin{equation}\label{k-range}
k_j\in\left\{-\frac{\pi(N-1)}{N},\dots,\frac{\pi(N-1)}{N},\pi\right\}\quad\quad {\rm for}\quad j=1,\dots,d,
\end{equation}


Notice, that (\ref{ak}) implies 
$\displaystyle A_{-\kgr}=A^\dagger_\kgr$ therefore $\displaystyle (A_{-\kgr}, A_{-\kgr})_0=(A_{\kgr}, A_{\kgr})_0$ by Prop.\ref{ilosk-prop}.
Writing   the inequality (\ref{bogol-1}) for $C^\dagger=C^\dagger_\kgr\,,\,A=A_\kgr$ and for $C^\dagger=C^\dagger_{-\kgr}\,,\,A=A_{-\kgr}$ and adding them, we obtain the inequality:
\begin{equation}\label{nier-pod-k}
\left|\om_0([C_{-\kgr}^\dagger, A_{-\kgr}])\right|^2 +\left|\om_0([C_{\kgr}^\dagger, A_{\kgr}])\right|^2\leq (A_\kgr,A_\kgr)_0 \,\om_0\left([[C_\kgr^\dagger, \ham_0],C_\kgr] +
[[C_{-\kgr}^\dagger, \ham_0],C_{-\kgr}]\right).
\end{equation}

\noindent
By the straightforward computation we have:
\begin{align*}
[C_\kgr^\dagger, A_\kgr]=& - \sum_{\xgr\ygr}e^{-i\kgr(\xgr-\ygr)}\left[e^{i\varphi_\xgr}(\partial_{\theta_\xgr}+ i \cot \theta_\xgr\partial _{\varphi_\xgr}), 
\cos \varphi_\ygr\sin\theta_\ygr P_l'(\cos\theta_\ygr)\right]=\\
=& \frac12\sum_\xgr e^{2 i \varphi_\xgr}\left\{2 \cos \theta_\xgr P_l'(\cos\theta_\xgr)-l(l+1)P_l(\cos\theta_\xgr)\right\} -\frac{l (l+1)}{2} \sum_\xgr P_l(\cos\theta_\xgr),
\end{align*}

The system consisting of spherical rotors possess the rotational symmetry i.e the hamiltonian (\ref{ham}) commutes with the unitary  operator $U_\varphi$ of rotation by an
angle $\varphi$ (in all sites simultanously):
\begin{equation*}
U_\varphi \ham U_\varphi^*= \ham,
\end{equation*}
where $\displaystyle \quad\quad(U_\varphi f)(\varphi_\xgr,\theta_\xgr, \varphi_\ygr,\theta_\ygr,\dots):=f(\varphi_\xgr-\varphi,\theta_\xgr, \varphi_\ygr-\varphi,\theta_\ygr,\dots).$\\
In particular  $\,U_\varphi P_0 U_\varphi^*= P_0.$ Since  $U_\varphi e^{i\varphi_\xgr} U_\varphi^* = e^{i(\varphi_\xgr-\varphi)} $, 
\begin{equation*}
\begin{split}
\om_0(e^{i\varphi_\xgr} F(\theta_\xgr))&=\tr(P_0e^{i\varphi_\xgr} F(\theta_\xgr)P_0)=\tr(U_\varphi P_0e^{i\varphi_\xgr} F(\theta_\xgr)P_0U_\varphi^*)
= \tr(P_0 U_\varphi e^{i\varphi_\xgr} F(\theta_\xgr)U_\varphi^*P_0 )=\\
&= e^{- i \varphi}\tr(P_0e^{i\varphi_\xgr} F(\theta_\xgr)P_0 )=
e^{- i \varphi} \om_0(e^{i\varphi_\xgr} F(\theta_\xgr)),
\end{split}
\end{equation*}
 for any (continuous) function $F$. Therefore      $ \,\om_0(e^{i\varphi_\xgr} F(\theta_\xgr))=0$.
In this way, using notation ({\ref{mh}) we get : 
$$\om_0( [C_\kgr^\dagger, A_\kgr ]) = -\frac{l(l+1)}{2}\, \om_0\left(\sum_\xgr P_l(\cos\theta_\xgr)\right)= -\frac{l(l+1)}{2} |\Lambda_N| \,m_N(h)=\om_0( [C_{-\kgr}^\dagger, A_{-\kgr} ]),$$
and  the LHS of (\ref{nier-pod-k}) can be expressed    in the form:
\begin{equation}\label{nier-LHS}
2 \left|\om_0([C_{\kgr}^\dagger, A_{\kgr}])\right|^2=\frac{l^2(l+1)^2}{2} |\Lambda_N|^2 \,m_N(h)^2
\end{equation}

Let us now compute the RHS of (\ref{nier-pod-k}). We start with $[C_\kgr^\dagger, \ekin]$:
%
$$[C_\kgr^\dagger, \ekin]=\frac{1}{2 I}\left[ \sum_\xgr e^{-i \kgr \xgr} L^+_\xgr, \sum_\ygr L^2_\ygr\right]=\frac{1}{2I} \sum_\xgr e^{-i \kgr \xgr} [L^+_\xgr,L^2_\xgr]=0$$
Notice that  $C_\kgr^\dagger$ is a first order differential operator and $\inter,\,V$ are operators of multiplication by smooth functions therefore 
$[C_\kgr^\dagger, \ham_0]=[C_\kgr^\dagger,V+\inter]$ is an  operator of multiplication by a smooth function as well, {\em so it is bounded} and the same is true for $[[C_\kgr^\dagger, \ham_0], C_\kgr]$.
Let us calculate:
%
\begin{equation*}
\begin{split}
[[C_\kgr^\dagger, \inter], C_\kgr]=&\sum_{\xgr, \ygr} e^{- i \kgr( \xgr-\ygr)}[[L_\xgr^+,\inter],  L_\ygr^-]=-h \sum_{\xgr, \ygr,\zgr} e^{- i \kgr( \xgr-\ygr)}[[L_\xgr^+,P_l(\cos\theta_\zgr)], L_\ygr^-]=\\
=&- h \sum_{\ygr,\zgr} e^{- i \kgr( \zgr-\ygr)}[- e^{i \varphi_\zgr}\sin \theta_\zgr P_l'(\cos\theta_\zgr), L_\ygr^-]= 
- h \sum_{\ygr} [ L_\ygr^-,  e^{i \varphi_\ygr}\sin \theta_\ygr P_l'(\cos\theta_\ygr)]=\\
=& h l (l+1) \sum_{\ygr} P_l(\cos \theta_\ygr),
\end{split}
\end{equation*}
where in the last equality, the Legendre equation (\ref{legendre-eq}) was used. It shows that  the  commutator does not depend on $\kgr$ and in particular
%
\begin{equation}\label{komut-int}
[[C_\kgr^\dagger, \inter],C_\kgr]=h l(l+1)\sum_\xgr P_l(\cos\theta_\xgr)= [[C_{-\kgr}^\dagger, \inter],C_{-\kgr}].
\end{equation}  
Let us  compute $ [[C_\kgr^\dagger, V],C_\kgr]$.
{
\begin{equation*}
[[C_\kgr^\dagger, V],C_\kgr]= \sum_{\xgr,\tgr}   e^{-\kgr(\xgr-\tgr)}[[L_\xgr^+,V],L_\tgr^-]=
\sum_{\xgr,\tgr,\ygr,\zgr}   e^{-\kgr(\xgr-\tgr)} J_{\ygr\zgr}\sum_{m=-l}^{l}(-1)^m[[L_\xgr^+,Y_l^m(\ngr_\ygr)Y_l^{-m}(\ngr_\zgr)],L_\tgr^-]
\end{equation*}
Now, using (\ref{comut-L-harm}) we get:
\begin{equation*}
\begin{split}
[L_\xgr^+,Y_l^m(\ngr_\ygr)Y_l^{-m}(\ngr_\zgr)]= & -\delta_{\xgr\ygr}\sqrt{(l-m)(l+m+1)}Y_l^{m+1}(\ngr_\ygr)Y_l^{-m}(\ngr_\zgr)+\\
&- \delta_{\xgr\zgr}\sqrt{(l+m)(l-m+1)}Y_l^{m}(\ngr_\ygr)Y_l^{-m+1}(\ngr_\zgr)
\end{split}
\end{equation*}
and by  (\ref{comut-L-harm}) again for $\displaystyle [L_\tgr^-, Y_l^{m+1}(\ngr_\ygr)Y_l^{-m}(\ngr_\zgr)]$ we finally obtain:
\begin{equation*}
\begin{split}
[[C_\kgr^\dagger, V],C_\kgr]=
\sum_{\ygr, \zgr} J_{\ygr \zgr} & \left( 2\,\left\{\cos \kgr (\ygr-\zgr) -1\right\} \sum_{m=-l}^l \left\{l(l+1)-m^2\right\}\,Y_l^{m}(\ngr_\ygr)\overline{Y_l^{m}(\ngr_\zgr)} +\right. \\
 & \left.- 2 i \sin \kgr(\ygr-\zgr) \sum_{m=-l}^l m Y_l^{m}(\ngr_\ygr) \overline{Y_l^{m}(\ngr_\zgr)} \right).
\end{split}
\end{equation*}
Replacing $\kgr$ by  $-\kgr$ in the  formula above and  adding both expressions we have
$$[[C_\kgr^\dagger, V],C_\kgr]+[[C_{-\kgr}^\dagger, V],C_{-\kgr}]=2 \sum_{\ygr \zgr}J_{\ygr \zgr}\,2 \left\{\cos \kgr(\ygr-\zgr) -1\right\}
\sum_{m=-l}^l \left\{l(l+1)-m^2\right\}\, Y_l^{m}(\ngr_\ygr)\,\overline{Y_l^{m}(\ngr_\zgr)}$$
and by (\ref{komut-int}):
%
%
\begin{equation}\nonumber 
\begin{split}
 \om_0\left([[C_\kgr^\dagger, \ham_0],C_\kgr]\right.&+\left.[[C_{-\kgr}^\dagger, \ham_0],C_{-\kgr}]\right)\, =\,  2 h l(l+1)|\Lambda_N|\,m_N(h)  + \\
&\quad+ 2 \sum_{\ygr \zgr}J_{\ygr \zgr }\,2\left\{\cos \kgr(\ygr-\zgr) -1\right\}\sum_{m=-l}^l \left\{l(l+1)-m^2\right\}\, \om_0\left(Y_l^{m}(\ngr_\ygr)\overline{Y_l^{m}(\ngr_\zgr)}\right).
\end{split}
\end{equation}
Notice that by  (\ref{nier-pod-k}):  
$$ \om_0\left([[C_\kgr^\dagger, \ham_0],C_\kgr]+[[C_{-\kgr}^\dagger, \ham_0],C_{-\kgr}]\right)= \left|  \om_0\left([[C_\kgr^\dagger, \ham_0],C_\kgr]+[[C_{-\kgr}^\dagger, H_0],C_{-\kgr}]\right)  \right|$$
and the estimation follows
\begin{equation}\label{nier-dod}
\begin{split}
\left|  \om_0\left( [[C_\kgr^\dagger, \ham_0],C_\kgr] \right.\right. &+ \left.\left. [[C_{-\kgr}^\dagger, \ham_0],C_{-\kgr}]\right)  \right| \leq  2 |h|  l(l+1)|\Lambda_N|\,|m_N(h)|+\\
+2 \sum_{\ygr \zgr}&|J_{\ygr \zgr }|\,2\left|\cos \kgr(\ygr-\zgr) -1\right|\sum_{m=-l}^l \left\{l(l+1)-m^2\right\}\, \left|\om_0\left(Y_l^{m}(\ngr_\ygr)\overline{Y_l^{m}(\ngr_\zgr)}\right)\right|.
\end{split}
\end{equation}
Let 
$$\Ygr(l):=max\{\sup|Y_l^m(\ngr)|\,,\,m=-l,\dots,l\}.$$ 
Then 
$$\displaystyle \left|\om_0\left(Y_l^{m}(\ngr_\ygr)\overline{Y_l^{m}(\ngr_\zgr)}\right)\right|\leq \Ygr(l)^2.$$
Now,  using the elementary inequality $\displaystyle 2 |\cos \kgr (\ygr-\zgr) -1|\leq |\kgr|^2|\ygr-\zgr|^2$ and the formula \linebreak[4]
$\displaystyle \sum_{m=-l}^l m^2=l(l+1)(2 l +1)/3$ we can estimate the RHS of (\ref{nier-dod}) by:
$$ 2 |h|  l(l+1)|\Lambda_N|\,|m_N(h)|+ \frac23 \Ygr(l)^2 2 l(l+1)(2 l +1) 
|\kgr|^2 \sum_{\ygr \zgr}|J_{\ygr \zgr }|\,|\ygr-\zgr|^2$$
By assumption (\ref{zal-J}):  $\displaystyle \sum_{\Z^d} J(|\xgr|) |\xgr|^2 \leq \cj$ therefore $\displaystyle \sum_{\ygr \zgr} J_{\ygr \zgr} |\ygr-\zgr|^2 \leq \cj |\Lambda_N| $ and we finally get:
\begin{equation}
\begin{split}
\left|  \om_0\left([[C_\kgr^\dagger, \ham_0],C_\kgr] + [[C_{-\kgr}^\dagger, \ham_0],C_{-\kgr}]\right)  \right| & \leq \\ 
2 l (l+1) |\Lambda_N|&\left[ |h| |m_N(h)|+ \frac23 (2l+1) \Ygr(l)^2 
\cj |\kgr|^2\right]
\end{split}
\end{equation}
Due to  this estimate and  the formula (\ref{nier-LHS}) we write the inequality (\ref{nier-pod-k}) as:
%
\begin{equation*}
\frac{l^2 (l+1)^2}{2}|\Lambda_N|^2 m_N(h)^2\leq (A_\kgr,A_\kgr)_0\, 2 l (l+1) |\Lambda_N|\left[ |h| |m_N(h)|+ \frac23 (2l+1) \Ygr(l)^2 
\cj |\kgr|^2\right]
\end{equation*}
or simply 
$$\frac{l (l +1)|\Lambda_N| m_N(h)^2}{4}\times \frac{1}{|h| |m_N(h)|+ \frac23 (2l+1) \Ygr(l)^2 
\cj |\kgr|^2}\leq (A_\kgr,A_\kgr)_0.$$
Summing over $\kgr$, where the range of summation is defined in (\ref{k-range}), we get:
\begin{equation}\label{nier-prawie-kon}
\frac{l (l +1)|\Lambda_N| m_N(h)^2}{4}\sum_{\kgr}\frac{1}{|h| |m_N(h)|+ \frac23 (2l+1) \Ygr(l)^2 
\cj |\kgr|^2}\leq \sum_{\kgr}(A_\kgr,A_\kgr)_0.
\end{equation}
Since $|m_N(h)|$ can be estimated as:
$$|m_N(h)|=\frac1{|\Lambda_N|}\left|\om_0\left(\sum_\xgr P_l(\cos\theta_\xgr)\right)\right|\leq \frac1{|\Lambda_N|}\sum_\xgr |\om_0(P_l(\cos\theta_\xgr))|\leq \sqrt{\frac{4 \pi}{2l+1}}\Ygr(l) $$
we can replace $|m_N(h)|$ in the denominator of (\ref{nier-prawie-kon}) by  $\sqrt{\frac{4 \pi}{2l+1}}\Ygr(l)$: 
$$\sqrt{\frac{2 l +1}{4 \pi}}\frac{l (l +1)|\Lambda_N| m_N(h)^2}{4 \Ygr(l) } \sum_{\kgr}\frac{1}{|h|+ \frac23 (2l+1)^{3/2}(4 \pi)^{-1/2} \Ygr(l) \cj |\kgr|^2}\leq \sum_{\kgr}(A_\kgr,A_\kgr)_0,$$
Finally we can write our inequality as:
\begin{equation}\label{final-spherical}
m_N(h)^2\,\frac{(2\pi)^d}{|\Lambda_N|}\sum_{\kgr}\frac{1}{|h| + K^2|\kgr|^2}\leq 4 \,(2\pi)^d\,\frac{ \Ygr(l)}{ l(l+1)}\sqrt{\frac{4\pi}{2 l +1}}\,F_N(h)
,
\end{equation}
where we have put $\displaystyle K:=\sqrt{\frac23 (2l+1)^{3/2}\ (4 \pi)^{-1/2}\Ygr(l) \cj} $ (note that $K$ does not depend on $N$) and 
\begin{equation}\label{def-F}
F_N(h):= \frac{1}{|\Lambda_N|^2  }\sum_{\kgr}(A_\kgr,A_\kgr)_0 
\end{equation}

\subsubsection{Planar rotors}
This time we  define  operators:
\begin{equation}
C^\dagger_\kgr:=\sum_{\xgr}e^{-\iota\kgr\xgr}\partial_{\varphi_\xgr}\quad,\quad A_\kgr:=- \sum_\xgr e^{\iota\kgr\xgr} \sin\varphi_\xgr
\end{equation}
As for spherical rotors we calculate their commutator
$$[C^\dagger_\kgr,A_\kgr]=- \sum_{\xgr\ygr}e^{-\iota\kgr(\xgr-\ygr)}[\partial_{\varphi_\xgr},\sin\varphi_\ygr]=-\sum_{\xgr}\cos\varphi_\xgr=[C^\dagger_{-\kgr},A_{-\kgr}].$$
and get   the LHS of the inequalitty (\ref{nier-pod-k}):
\begin{equation}\label{LHS-planar}
2 |\om_0([C^\dagger_\kgr,A_\kgr])|^2 =2 |\Lambda_N|^2 m_N(h)^2\,, \,\,
\end{equation}
where, in this case,  the magnetization is defined by (\ref{mh-planar}).

Now, for the RHS of  (\ref{nier-pod-k}):
$$[C^\dagger_\kgr,\ekin]=0$$
$$[[C^\dagger_\kgr,\inter], C_\kgr]= h \sum_{\xgr\ygr\zgr} e^{-\iota\kgr(\xgr-\zgr)}[[\partial_{\varphi_\xgr},\cos\varphi_\ygr],\partial_{\varphi_\zgr}]=
h \sum_{\xgr\zgr} e^{-\iota\kgr(\xgr-\zgr)}[-\sin\varphi_\xgr,\partial_{\varphi_\zgr}]=h \sum_{\xgr} \cos\varphi_\xgr$$
$$[[C^\dagger_\kgr, V],C_\kgr]=\sum_{\xgr\ygr\zgr\tgr} J_{\ygr\zgr} e^{-\iota\kgr(\xgr-\tgr)}[[\partial_{\varphi_\xgr},\cos\varphi_\ygr\cos\varphi_\zgr+\sin\varphi_\ygr\sin\varphi_\zgr],\partial_{\varphi_\tgr}]$$
and, after routine calculations, we get:
\begin{equation*}\begin{split}
[[C^\dagger_\kgr, V],C_\kgr] & =2 \sum_{\xgr\ygr}J_{\xgr\ygr}\,(1-e^{-\iota\kgr(\xgr-\ygr)})\,(\cos\varphi_\xgr\cos\varphi_\ygr+\sin\varphi_\xgr\sin\varphi_\ygr)=\\
&= 2 \sum_{\xgr\ygr}J_{\xgr\ygr}\,(1-\cos\kgr(\xgr-\ygr))(\cos\varphi_\xgr\cos\varphi_\ygr+\sin\varphi_\xgr\sin\varphi_\ygr)
\end{split}
\end{equation*}
(we have used the symmetry $J_{\xgr\ygr}=J_{\ygr\xgr}$). As for spherical rotors we obtain:
\begin{equation*}
\begin{split}
|\om_0([[C_\kgr^\dagger, \ham_0],C_\kgr]+[[C_{-\kgr}^\dagger, \ham_0],C_{-\kgr}])|& \leq \,  2 |h| |\Lambda_N| |m_N(h)|+\\& 4 \sum_{\xgr\ygr}|J_{\xgr\ygr}|\, |1-\cos\kgr(\xgr-\ygr)|
|\om_0(\cos\varphi_\xgr\cos\varphi_\ygr+\sin\varphi_\xgr\sin\varphi_\ygr)|
\end{split}\end{equation*}
Now, using inequalities  $\displaystyle\, 2|1-\cos\kgr(\xgr-\ygr)|\leq |\kgr|^2 |\xgr-\ygr|^2$, 
$\displaystyle \sum_{\xgr \ygr} J_{\xgr \ygr} |\xgr-\ygr|^2 \leq \cj |\Lambda_N|\,,$ and  $\displaystyle |\om_0(\cos\varphi_\xgr\cos\varphi_\ygr+\sin\varphi_\xgr\sin\varphi_\ygr)|\leq  1$
we have
$$|\om_0([[C_\kgr^\dagger, H],C_\kgr]+[[C_{-\kgr}^\dagger, H],C_{-\kgr}])|\leq 2|\Lambda_N|\left(|h| |m_N(h)|  + \cj |\kgr|^2  
\right);$$
and  using (\ref{nier-pod-k}) and  (\ref{LHS-planar}):
$$2 |\Lambda_N|^2 |m_N(h)|^2\leq 2 (A_\kgr,A_\kgr)_0 |\Lambda_N|\left(|h| |m_N(h)|  + \cj |\kgr|^2 
   \right),$$
which, in turn (note the  obvious estimate $|m_N(h)|\leq 1 $), can be written as:
$$m_N(h)^2 \frac{(2 \pi)^d}{|\Lambda_N|}\times\frac1{|h|  + \cj |\kgr|^2 }\leq  \frac{(2\pi)^d}{|\Lambda_N|^2} (A_\kgr,A_\kgr)_0  $$
Finally, performing summation over $\kgr$ (as for spherical rotors) and  using (\ref{def-F}) :
\begin{equation}\label{final-planar}
m_N(h)^2 \frac{(2 \pi)^d}{|\Lambda_N|}\,\sum_{\kgr}\frac1{|h|  + K^2 |\kgr|^2 }\leq (2\pi)^d\,F_N(h),
\end{equation}
where, this time, we have put $K:=\sqrt{\cj}$. This way  we have the same expression(formally)  as (\ref{final-spherical}).

\subsection{Conditions for vanishing magnetization} In dimensions $d=1$ or $d=2$, the boundeness of the RHS of (\ref{final-planar}) (or (\ref{final-spherical}))  
forces magnetization to vanish (in $h\rightarrow 0$) limit. More precisely :

\begin{prop} Let $d=1$ or $d=2$. If for some $\delta>0$:
$$\sup\,\{F_N(h): 0<|h|\leq \delta\,,N\in \N \} < \infty, $$ 
then  
\begin{equation}
\displaystyle \,\,\lim_{h\rightarrow 0}\limsup_{N\rightarrow \infty}|m_N(h)|=0.
\label{Lim_hto0}
\end{equation}
\end{prop}
{\em Proof:} Let us denote, for $h\neq 0$, 
\begin{equation*}
S_N(h):=\frac{(2\pi)^d}{|\Lambda_N|}\sum_{\kgr}\frac{1}{|h| + K^2|\kgr|^2}> 0\,,\quad
I_d(h):= \int_{[-\pi,\pi]^d}\frac {d \kgr}{|h| + K^2|\kgr|^2}.
\end{equation*}
Then it is easy to see that $\displaystyle \lim_{N\rightarrow \infty} S_N(h)= I_d(h)$ and for $d=1,2$   $\displaystyle \lim_{h\rightarrow 0}I_d(h)=\infty$.
Take $h\neq 0$ such that $|h|\leq \delta$.
By the assumption, for any $N$  we can write inequalities  (\ref{final-spherical}) and (\ref{final-planar}) as 
$$|m_N(h)|\leq \sqrt{\frac{M}{S_N(h)}}$$
for some positive constant $M <\infty$.
Let us define (for $h\neq 0$) a sequence $s_N(h):=\inf\{S_{N'}(h)\,,\, N'\geq N\}$. 
Since $\displaystyle \lim_{N\rightarrow \infty} s_N(h)=I_d(h),$   $s_N(h)>0$ for sufficiently large $N$, $N\geq N_0$. Now, for $N\geq N_0$:

$$|m_N(h)|\leq \sqrt{\frac{M}{S_N(h)}}\leq \sqrt{\frac{M}{s_{N_0}(h)}}$$
therefore
$$\sup_{N\geq N_0}|m_N(h)|\leq\sqrt{\frac{M}{s_{N_0}(h)}}$$
and
$$
\lim_{N_0\rightarrow\infty}\sup_{N\geq N_0} |m_N(h)|\leq \sqrt{\frac{M}{I_d(h)}}
$$
Now the assertion is clear.
\dowl

\noindent
{\bf Remark.} 
The inequality (\ref{Lim_hto0}) is weaker than the standard ones. For instance, in $d=1$ Heisenberg models,
one obtains more explicite estimation \cite{MW}: 
\[
m(h)\leq {\rm Const}\, h^{1/3} T^{-2/3}.
\]
However, in such estimations, certain somewhat hidden fact is used when the thermodynamic limit is taken. Namely,
 the existence of free energy and magnetization in
thermodynamic limit is assumed. It is in fact true \cite{Ruelle}. However, we didn't use this fact and the estimation
(\ref{Lim_hto0}) is weaker than the standard ones, but sufficient to show the absence
 of the spontaneous magnetization at zero
magnetic field.

We describe the sufficient condition to ensure the boundedness of $F_N(h)$.
Let us  {\em assume} the gap in energy spectrum:
\begin{equation*}
\inf_{h} \inf_{\Lambda_N}\{E_1-E_0\}=:\Delta >0.
\end{equation*}
In this situation, the computation in the proof of Proposition \ref{prop-bogol-1} applied to $A_\kgr$ leads  to the estimate (since $A_\kgr$ is normal):
$$(A_\kgr,A_\kgr)_0\leq \frac{2}{\Delta} \omega_0(A_\kgr A^*_\kgr)$$
Using the definition (\ref{ak}) of $A_\kgr$  we compute for {\em spherical rotors}:
\begin{equation*}
\begin{split}
\sum_\kgr A^*_\kgr A_\kgr & =\sum_\kgr\sum_{\xgr\ygr}e^{i\kgr(\xgr-\ygr)} 
\cos\varphi_\xgr \sin\theta_\xgr  P_l'(\cos\theta_\xgr)\cos\varphi_\ygr \sin\theta_\ygr  P_l'(\cos\theta_\ygr)=\\
&=|\Lambda_N| \sum_{\xgr}\cos^2\varphi_\xgr \sin^2\theta_\xgr  (P_l'(\cos\theta_\xgr))^2\leq |\Lambda_N|^2 \tilde{Y}(l), 
\end{split}\end{equation*}
where $\tilde{Y}(l):=\sup \left\{(P_l'(\cos\theta))^2\,\sin^2\theta\right\}$ and we have used $\sum_\kgr e^{i\kgr(\xgr-\ygr)} =|\Lambda_N|\delta(\xgr-\ygr)$. \\
Thus $\displaystyle F_N(h)\leq\tilde{Y}(l)$ and the upper bound  does not depend  on  $\Lambda_N$ and $h$.

For {\em planar rotors}  we calculate:
\begin{equation*}
\begin{split}
\sum_\kgr A^*_\kgr A_\kgr & =\sum_\kgr\sum_{\xgr\ygr}e^{i\kgr(\xgr-\ygr)} \sin\varphi_\xgr\sin\varphi_\ygr=|\Lambda_N|\sum_\xgr\sin^2\varphi_\xgr\leq |\Lambda_N|^2
\end{split}
\end{equation*}
and, in this case, $\displaystyle F_N(h)\leq1$.


\subsection{Arguments for existence of energy gap in one-dimensional system of planar rotors}

In this  subsection we  argue that an energy gap should be present in the system of rotors if their 
coupling is sufficiently small. Consequently, such a system should not be  ordered also in the
ground state.

More precisely, we analyse one-dimensional system of planar  rotors with nearest neighbour interactions.
We slightly redefine  (by a simple rescaling)
the Hamiltonian written in the  Subsec. 3.1: 
\begin{equation}\label{Ham1d}
\ham= \ekin + V,
\end{equation}
where
\[
\ekin=
-\sum_\xgr \frac{\partial^2}{\partial \varphi_\xgr^2},
\]
\[
V=
\lambda \sum_{\langle\xgr,\ygr\rangle}  \cos(\varphi_\xgr -\varphi_\ygr) - h \sum_\xgr \cos\varphi_\xgr
\]
The total number of rotors is  $N$, and we impose periodic boundary condition.
We will analyse energy levels of the system in the framework of perturbation theory, assuming that the 
coupling constant $\lambda$ characterizing interaction between rotors is sufficiently small. 

First, let us consider the rotor system in zero magnetic field, i.e. the case $h=0$.
The unperturbed system (i.e. for $\lambda=0$) is described by the Hamiltonian  $\ekin$;
this is a system of $N$ non-interacting rotors. For a single rotor, the eigenfunctions $\psi_k$
and corresponding eigenenergies $E_k$ are
\begin{equation}
\psi_k = e^{i k \phi},\quad E_k = k^2, \quad k\in\mathbb{Z}
\label{ExpBasis}
\end{equation}
Sometimes it is more convenient to use the real 'trigonometric' basis:
\begin{equation}
s_k = \sin k \phi, \quad k\in \mathbb{N};\qquad
c_k = \cos k \phi, \quad k\in \mathbb{N} \cup \{0\}.
\label{TrigBasis}
\end{equation}
(eigenenergies of both $c_k$ and $s_k$ are $E_k=k^2$). For $N$ rotors,
the eigenfunction is a tensor product of $N$ eigenfunctions on every site:
\[
\Psi \equiv \Psi_{k_1, \dots, k_N} 
=
\psi_{1,k_1} \otimes \psi_{2,k_2} \otimes \dots  \otimes \psi_{N,k_N}
\] 
(for $\psi_{m,k_m}$, the first index $m$ is the site index, and $k_m$ is the number of eigenfunction on
the site $m$). For trigonometric basis we have analogous expression. The eigenenergy of 
the $\Psi$ function is
\[
E_\Psi = k_1^2 + k_2^2+\dots+k_N^2
\]

The  \em unique} ground state $\Psi_0$ corresponds to $k_1=k_2=\dots=k_N=0$  and energy $E_0=0$.

The first excited state is obtained when one of rotors (say, $j-$th) is in the first excited state 
(i.e. $k_j = \pm 1$) whereas remaining rotors are in ground states. It is clear that the first excited
state of the system has degeneracy equal to $2N$. Its eigenenergy $E_1$ is:   $E_1=1$. Notice that the energy gap
$E_1-E_0 =1$ and that {\em it is independent of $N$}.

Now, let us calculate first-order correction to the ground state and the first excited state. We apply the
standard perturbation theory, see for instance \cite{Schiff} or \cite{RS4}.

 {\em The ground state.} The first-order correction  $\Delta E^{(1)}_0$ is zero:
\[
\Delta E^{(1)}_0 = \langle \Psi_0 | V | \Psi_0\rangle=0
\]

{\em The first excited state.} The unperturbed state has $2N-$fold degeneracy, so the degenerate perturbation
theory must be applied. Let us enumerate eigenfunctions  of unperturbed system, which correspond to energy $E_1$ as:
 $\Psi_{1,1}, \Psi_{1,2},\dots, \Psi_{1,j}, \dots,\Psi_{1,2N}$. Define the $V_{jm}$ matrix as:
\[
V_{jm} = \langle \Psi_{1,j}|V| \Psi_{1,m} \rangle
\] 
The $1-$st order corrections to the first excited state are eigenvalues of the  $V_{jm}$ matrix.
We will apply the trigonometric basis (\ref{TrigBasis}).
By a straightforward calculation, it turns out that macierz $V_{jm}$ factorizes into two identical matrices.
Due to cyclic boundary conditions, they are {\em cyclic} matrices: $V_{j,j+1}=V_{j,j-1}=\lambda$,
$V_{1,N}=V_{N,1}=\lambda$; other matrix elements are zero. Their eigenvalues $\lambda_n$ are:
\[
\lambda_n=2\lambda \cos \frac{2\pi n}{N}. 
\]
The lowest value among first-order correction is: 
\[
E^{(1)}_{1,min} = 1-2\lambda
\]
and this is also the value of energy gap in the first order perturbation theory. Notice that this is {\em independent of} $N$.

{\em Remark.} The reasoning above can be extended to the following situations:
\begin{itemize}
\item In considerations above we assumed that the magnetic field is zero. For non-zero magnetic fields, the unperturbed Hamiltonian
corresponds to uncoupled rotors in periodic potential. Eigenvalues and eigenvectors of such a system are explicitely known and 
are expressible by {\em Mathieu function} \cite{WhittakerWatson}. They are much more involved, but the result is the same as in the case
of zero magnetic field: {\em For sufficiently small magnetic field, the system of rotors exhibits the energy gap above the ground  state
 in the first order of perturbation theory. The value of gap is independent of $N$.}
\item The interaction of rotors can be arbitrary finite range, translationally invariant but small; under these assumptions, the energy
gap is also present in the first order of perturbation theory. 
\item These results hold also for {\em two-dimensional} lattice, by a straightforward extension of arguments above.
\end{itemize}
\section{Summary, perspectives}
\label{sec:SummPer}
In the paper, we have formulated and proved the Bogolyubov inequality for operators at zero temperature.
So far this inequality has been known for matrices, and we were able to extend it to operators. We have also 
applied this inequality to the system of interacting rotors. We have shown that if the dimension of the
lattice is 1 or 2,  the interaction decreases  sufficiently fast  with  a distance (cf. (\ref{zal-J})) and  there is an energy gap over the ground state, then the spontaneous
magnetization in the ground state is zero, i.e. there is no LRO in the system.
 
 We present also heuristic arguments suggesting that one- and two-dimensional system of interacting rotors
 has the energy gap independent of the system size
 if the interaction is sufficiently small. This would imply the lack of ordering in the
 ground state of such rotor system. The argument is based on perturbation theory. Unfortunately we were not able
 to proceed further with perturbation theory -- the calculations become hopelessly complicated in further
 orders of perturbation theory. This way, the rigorous proof of disorder in the ground state of weakly interacting
 rotors is still an open problem. 
Perhaps other methods could be more adequate. One of them is {\em stochastic analysis} used in the anharmonic crystal
model problems \cite{VerbZagr}, \cite{ACM}. Another potentially applicable method could be the
rigorous renormalization group \cite{Mastropietro}.
 In the case where qualitative
properties of unperturbed system are suspected to be not changed under switching on the perturbation,
one can hope to show rigorously the convergence of RG perturbation series.

To conclude, let us mention some other open (to our best knowledge) problems, which seem for us to be very interesting.

For {\em large} couplings between rotors, the ground state is ordered (for $d=2$ and
nearest-neighbour ferromagnetic couplings) \cite{W1}, \cite{WPS1}. 
For {\em small} couplings, presumably there is no ordering in the ground state. So, for some intermediate 
value of coupling, we conjecture that  the {\em quantum critical point} should appear. It would be very interesting to verify 
such a conjecture, and if it is true, to examine the nature of this critical point.
\section{Appendix}
\label{sec:App}
Here we collect definitions and formulae for spherical harmonics, Legendres' polynomials and angular momentum operators we use:
$$P_l(x):=\frac{(-1)^l}{2^l l!}\frac {d^l}{dx^l}(1-x^2)^l$$
\begin{equation}\label{legendre-eq}
(1-\cos^2\theta)P_l''(\cos\theta)-2 \cos\theta P_l'(\cos\theta)+ l (l+1) P_l(\cos\theta)=0;
\end{equation}
$$P_l^m(x):=(-1)^m(1-x^2)^{m/2}\frac {d^m}{dx^m}P_l(x)\,,\quad P_l^{-m}(x):=(-1)^m\frac{(l-m)!}{(l+m)!} P_l^{m}(x)\,,\quad m=0,1,\dots, l$$
$$Y_l^m(\theta,\varphi):=(-1)^m\sqrt{\frac{(2 l + 1)(l-m)!}{4 \pi (l+m)!}}P_l^{m}(\cos\theta) e^{i m \varphi}=:C_l^m P_l^{|m|}(\cos\theta) e^{i m \varphi}\quad,\quad m=-l,\dots , l$$
$$C_l^m:=\left\{\begin{array}{lr} (-1)^m\sqrt{\frac{(2 l + 1)(l-m)!}{4 \pi (l+m)!}}&m\geq 0\\
\sqrt{\frac{(2 l + 1)(l-m)!}{4 \pi (l+m)!}}& m < 0\end{array}\right.$$
$$\overline{Y_l^{m}}=(-1)^m Y_l^{-m}$$ 
$$L^+:=e^{i \varphi}(\partial_\theta+ i \cot \theta \partial_\varphi)\quad,\quad L^-:=(L^+)^*=e^{- i \varphi}(- \partial_\theta+ i \cot \theta \partial_\varphi)\quad,\quad L^z:=-i\partial_\varphi$$
$$[L^+,L^-]=2L^z\quad,\quad [L^z,L^+]=L^+\quad,\quad[L^z,L^-]=- L^-$$
$$L^2=-(\partial^2_\theta +\cot \theta \partial_\theta +\frac1{\sin^2\theta}\partial^2_\varphi)=L^+L^-+(L^z)^2-L^z$$
\begin{equation}\label{comut-L-harm}
[L^+, Y_l^m]=-\sqrt{(l-m)(l+m+1)} Y_l^{m+1},\,[L^-, Y_l^m]=-\sqrt{(l+m)(l-m+1)} Y_l^{m-1}
\end{equation}


\end{document}